\documentclass[11pt]{article}
\usepackage{DGfest,epsfig}
\usepackage{times}
\input colordvi 

\def\be{\begin{equation}}
\def\ee{\end{equation}}
\def\bea{\begin{eqnarray}}
\def\eea{\end{eqnarray}}

\begin{document}
\baselineskip 11.5pt
\vspace*{-25mm}
\begin{flushright}
     \small HU--EP--06/02\\
     January 2006
\end{flushright}
\vspace{6mm}
\title{\Large Cooling, smearing and Dirac eigenmodes - \\
\vskip0mm
A comparison of filtering methods in lattice gauge theory
\footnote{Invited contribution 
to {\sl Sense of Beauty in Physics}, 
  Festschrift in honor of
  Adriano Di Giacomos' 70-th birthday.}}

\author{ Christof Gattringer }

\address{Institut f\"ur Physik, FB Theoretische Physik, Unversit\"at Graz\\ 
8010 Graz, Austria }

\author{ E.-M.\ Ilgenfritz }

\address{ Institut f\"ur Physik, Humboldt Universit\"at zu Berlin\\ 
12489 Berlin, Germany }

\author{ Stefan Solbrig }

\address{ Institut f\"ur Physik,
Universit\"at Regensburg\\ 
93040 Regensburg, Germany }

\maketitle\abstracts{ Starting from thermalized quenched SU(2) 
configurations we apply cooling or iterated smearing, respectively,
to produce sequences
of gauge configurations with less and less fluctuations. We compute the low
lying spectrum and eigenmodes of the lattice Dirac operator and compare them
for the two types of smoothing. Many characteristic properties
of the eigensystem remain invariant for all configurations in our sequences. 
We also find that cooling and smearing produce surprisingly similar results. 
Both observations could be indications that the two filtering methods do not
drastically alter the long range structures in the gauge field.}

\section{Introduction}

An appealing feature of numerical lattice QCD is that it allows for 
a direct access to gluon field configurations as they appear in the fully
quantized path integral. The motivation for their analysis is the hope to 
identify the key mechanisms giving
rise to the characteristic features of confinement and chiral symmetry
breaking. Unfortunately, a naive look at, e.g., the action density reveals only
quantum fluctuations. However, it is widely believed that underneath these UV
fluctuations long range structures are hidden which might be stabilized by
topology.

When analyzing these infrared structures a filter for removing the quantum
fluctuations is necessary. Three main approaches can be found in the
literature. Cooling is essentially a Monte Carlo 
update which accepts only changes
that lower the action, thus finally driving the configuration into a classical
solution. Smearing is an operation which combines neighboring 
gauge variables to average out short distance fluctuations. Repeated
application thus also gives rise to smoother and smoother gluon configurations.
Both, cooling and smearing do actually change the gauge field to extract its
IR content. Thus an obvious criticism of these two approaches is the
uncertainty whether the filtering does not also destroy or at least considerably change
the long range structures one is interested in.

An alternative which has become widely popular in the last few years
\cite{diracfiltering} is the analysis
of low lying eigenmodes of the lattice Dirac operator and very recently also
of the covariant Laplace operator \cite{laplacefiltering}. 
It is expected that the low lying eigenmodes
couple mainly to the infrared structures of the gauge field. Obviously the
method leaves the gauge field untouched while exploring the long
range structures. It is considered to be a particularly ``physical'' filter
since the IR content of the gluon field is ``seen through the eyes of the
quarks''. 

However, a big disadvantage of eigenmode filtering is that purely 
gluonic quantities such as the field strength, the action- and charge
densities, or the Polyakov
loop have to be reconstructed in a non-trivial way 
\cite{laplacefiltering,selfdual}, or are not accessible
at all. Since these gluonic observables provide important information 
for many physical questions, it is desirable to have 
an alternative to the fermionic filter, which at the same time guarantees 
that the long range structures are not altered.  

In this little study we address in a new way 
the question whether cooling \cite{rapidcooling,slowcooling} or smearing
\cite{smearing} alter the infrared content of the gauge field: We use
observables based on the low lying 
Dirac eigenmodes and analyze how they change under 
cooling or repeated smearing. If such fermionic observables change 
considerably in a sequence of smoother and smoother configurations, this is a
strong indication that the IR structures are altered by the smoothing
method applied. Here we present evidence that for many steps of smoothing
(cooling or smearing) the fermionic observables remain essentially
invariant. Furthermore we find that the two smoothing methods we use,
cooling and smearing, lead to very similar results.  

\section{Technicalities}

Our study is based on quenched SU(2) configurations generated with the
L\"uscher-Weisz gauge action \cite{Luweact} 
at $\beta = 1.95$ on lattices of size $16^4$. We use periodic boundary
conditions for the gauge fields. As outlined in
the introduction, we generate from the original, thermalized configurations
sequences of smoother and smoother configurations by applying either cooling
or smearing. 

In our cooling procedure we run the same single link Metropolis update 
used for generating the
thermalized configurations but omit the conditional acceptance step for the
case where the offered gauge configuration leads to an increased action. The
offer is chosen relatively close to the old link variable (see also restricted
cooling \cite{slowcooling}). This implies that our cooling is very modest and
we can produce sequences of cooled configurations with only small changes in
each sweep. The number of cooling sweeps was chosen such that the average
plaquette matches the value obtained by smearing (compare Table 1). 

The 4-dimensional smearing of our Monte Carlo gauge field configurations
has been performed in the standard way  of sequential substitution sweeps:
\begin{eqnarray}
U^{(j)}_{x,\mu} \to U^{(j+1)}_{x,\mu} & = &
\mathrm{Proj}_{SU(2)} \bigg[ \; \alpha \, U^{(j)}_{x,\mu} \\
& + &
\gamma \sum_{\nu\ne\mu}
\left( U^{(j)}_{x,\nu} U^{(j)}_{x+\hat{\nu},\mu} U^{(j) \dagger}_{x+\hat{\mu},\nu}
+ U^{(j) \dagger}_{x-\hat{\nu},\nu} U^{(j)}_{x-\hat{\nu},\mu}
U^{(j)}_{x-\hat{\nu}+\hat{\mu},\nu} \right) \bigg] \; ,
\nonumber
\end{eqnarray}
with $\alpha= 0.55$ and $\gamma=(1-\alpha)/6= 0.075$. Obviously, this standard
smearing is not adapted to any specific action that has been used for the
generation of the Monte Carlo ensemble. 

\begin{table}[t]
\begin{center}
\begin{tabular}{c|c|c|c}
\multicolumn{2}{c}{cooling} &
\multicolumn{2}{c}{smearing} \\
\hline
sweeps & $\langle U_p \rangle$ & sweeps & $\langle U_p \rangle$ \\
\hline
0    & 0.695165 &  0 & 0.695165 \\
8    & 0.884064 &  1 & 0.885551 \\
19   & 0.948482 &  2 & 0.949688 \\
38   & 0.973993 &  3 & 0.974081 \\
72   & 0.984799 &  4 & 0.984740 \\
140  & 0.990499 &  5 & 0.990010 \\
220  & 0.992822 &  6 & 0.992907 \\
360  & 0.994559 &  7 & 0.994647 \\
580  & 0.995730 &  8 & 0.995771 \\
900  & 0.996520 &  9 & 0.996541 \\
1360 & 0.997089 & 10 & 0.997093 
\end{tabular}
\caption{
Comparison of key numbers for a cooling and a smearing sequence, both
starting from the same thermalized configuration. We display the number of 
cooling- and smearing steps respectively and the corresponding value of the
average plaquette. The number of cooling sweeps was chosen such that the
cooled configuration matches the action of the corresponding smeared
configuration as close as possible. \hfill $^{\;}$}
\end{center}
\end{table}

We consider a total of 10 thermalized configurations and apply to each of
them 10 sweeps of smearing, storing the configurations after each sweep. 
Starting from the same 10 thermalized configurations we also apply suitable numbers of 
cooling sweeps such that the action is as close as possible to the
corresponding smeared configurations.
Thus we end up with two times 10 sequences of configurations each
consisting of the original gauge field and 10 increasingly smoothed
configurations.  
For these sequences of cooled, respectively smeared configurations we calculate
the lowest 30 eigenvalues and eigenvectors of the lattice Dirac operator with 
the Arnoldi method \cite{arnoldi}. We use the chirally improved Dirac 
operator \cite{chirimp}, an approximate solution of the 
Ginsparg-Wilson equation \cite{giwi}, which is the optimal 
implementation of chiral symmetry 
on the lattice. 

Of the 10 thermalized configurations we analyze, 2 have topological charge 
$Q = 0$, 6 have $|Q| = 1$ and the remaining 2 are in the sector $|Q| = 2$,
with the topological charge determined via the index theorem through 
the number of left- or right-handed zero modes.

We compute the eigensystem for both, periodic and
anti-periodic temporal boundary conditions, leaving the spatial boundary
conditions periodic. The motivation for two different temporal boundary
conditions is that 
for finite temperature it is known \cite{kvbdirac}
that the zero modes are localized on monopole-like constituents of 
so-called Kraan-van Baal calorons \cite{kvb} which may be located at different
space-time positions. Also for configurations on
the torus the corresponding hopping of the zero modes 
has been observed for gauge group SU(2) and 
SU(3) \cite{sunhop}. Here we want to probe the sequences of
configurations with the eigensystem of the Dirac operator, and the hopping of 
the zero modes under a change of boundary conditions is highly welcome 
as an additional signature. 

The base quantity for our fermionic observables is the scalar density
\begin{equation}
\rho(x) \; = \; \sum_{c,\alpha} | v_{c,\alpha}(x) |^2 \; ,
\end{equation}
obtained by locally summing the absolute square for each entry of a Dirac
eigenvector $v$ over its color and Dirac indices $c$ and $\alpha$.
In order to determine the location of a mode we use the maximum
of $\rho(x)$. Using 3-d plots of $\rho(x)$ over slices of the lattice 
(see Fig.\ \ref{slices}) we can further analyze the eigenmodes.
A quantity which condenses the localization properties of eigenmodes into a
single number is the inverse participation ratio 
\begin{equation}
IPR \; = \; V \sum_x \rho(x)^2 \; .
\end{equation}
Using the fact that $\sum_x \rho(x) = 1$ (our eigenvectors are normalized to
1), it is easy to see that the inverse participation ratio reaches its maximum
value $IPR = V$ for a density $\rho$ which is 1 for a single site and 0 everywhere
else, while for a completely spread out eigenmode ($\rho(x) \equiv 1/V$) 
one finds $IPR = 1$. 

\section{Results}

\begin{figure}[t]
\centering
\includegraphics[width=12.5cm,clip]{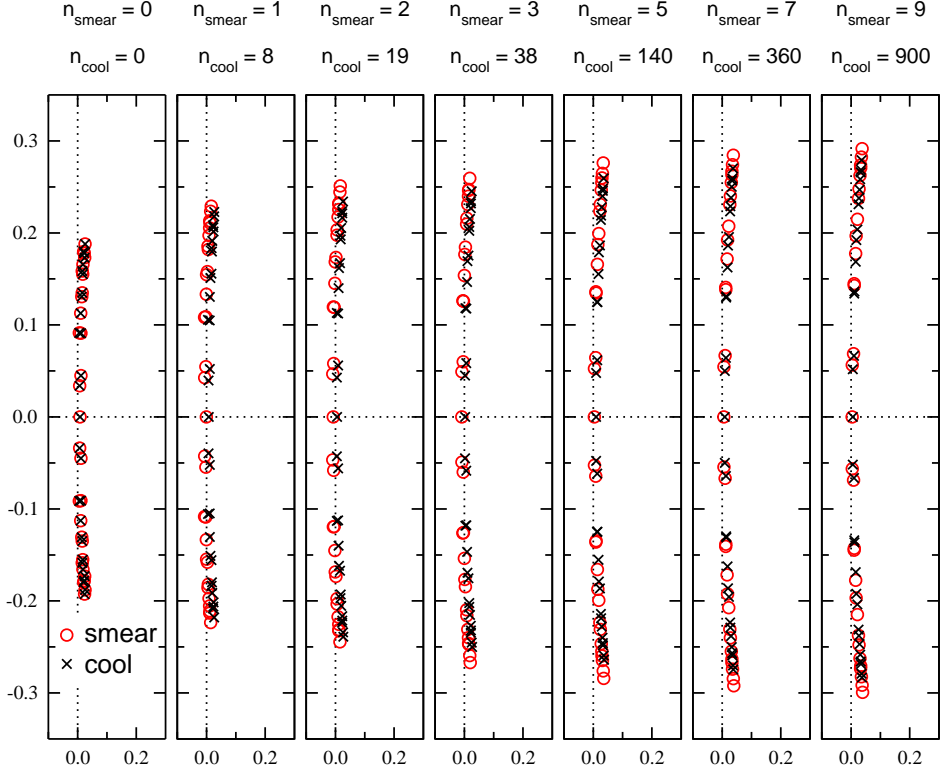}
\caption{Spectra of the Dirac operator in the complex plane. The circles 
  represent eigenvalues for the sequence of configurations obtained by
  smearing, the crosses are used for the cooled configurations. In each spectrum
  the 30 smallest eigenvalues for periodic boundary conditions are shown.
\hfill $^{\;}$}
\label{speccompare}
\end{figure} 

\subsection{Eigenvalues and topological charge}
We begin the presentation of our results with the discussion of the
eigenvalues and their change under the two smoothing procedures. Fig.\ 
\ref{speccompare} shows the 30 lowest lying eigenvalues of the Dirac operator 
in the complex plane for one pair of our sequences. The circles are used for
the smeared configurations, while the crosses represent the results for
cooling. The left-most spectrum is for the original, thermalized configuration
and as one moves to the right the amount of smoothing increases. The
corresponding numbers of cooling and smearing steps respectively are denoted
at the top of each individual plot.   

A feature, which is obvious from the plots, is that when smoothing the 
configurations, the spectrum becomes less dense and the eigenvalue density is 
reduced. Smearing stretches the spectrum slightly stronger, thus reducing the
eigenvalue density by a few percent more than cooling. Otherwise the 
eigenvalues, in particular the lowest lying ones, behave almost identically 
under cooling and smearing, and the symbols lie nearly on top of each other.
Also for the higher lying eigenvalues the characteristic patterns for their
grouping are kept by both smoothing methods. 

The configuration we have used for Fig.\ \ref{speccompare} has a single, 
left-handed zero mode and through the index theorem thus can be seen to
have topological charge $Q = 1$. The existence of this zero mode is not
altered by the two smoothing procedures. This observation holds for all our 10
pairs of sequences: The topological charge as determined by the index theorem
did never change under the cooling or smearing we applied. We should however
remark that much longer sequences of cooling or smearing will certainly destroy
the topological charge and thus lead to a vanishing of the corresponding zero
mode. 

We remark that for staggered fermions it was shown \cite{DuHo} in a similar way that
under smearing the spectral properties become more transparent then. A version
of the index theorem, suitable for staggered fermions, can be established.

\begin{figure}[p]
\centering
\includegraphics[width=14.2cm,clip]{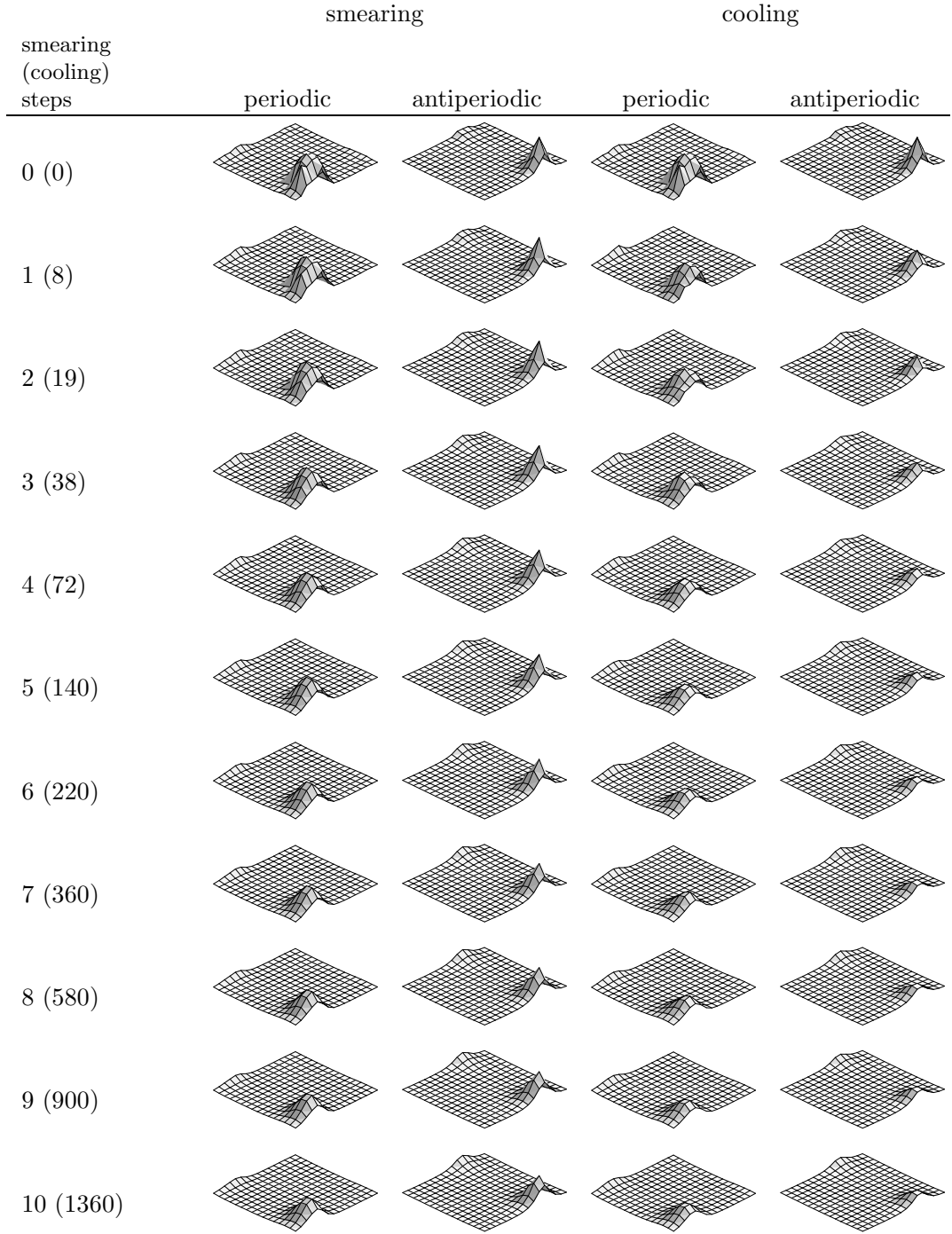}
\vskip3mm
\caption{Scalar density of a zero mode for periodic and anti-periodic
  boundary conditions. The two columns on the left-hand side are for the
  sequence of configurations generated with smearing, while the two columns on
  the right-hand side are for the corresponding sequence from cooling. For both, periodic
  and anti-periodic boundary conditions we show the $xy$-slice through the
  respective maximum. \hfill $^{\;}$}
\label{slices}
\end{figure} 

\subsection{Scalar density and the role of the boundary condition}

Let us now address the behavior of the eigenmodes under smoothing. In
particular we here focus on the 6 configurations with exactly one eigenvalue zero  
and present results for the corresponding zero modes. As announced, our basic observable
is the scalar density $\rho(x)$, 
and in a first step we look for the lattice site where
it assumes a maximum. Subsequently we study the scalar density over slices
of the lattice cut through the maximum. 
For our 4-dimensional lattice we have 6 different slices and we
typically find that the lumps in the scalar density are localized in all 4
directions. Thus defining the location 
of an infrared lump through the position of its 
maximum is a meaningful procedure. 

In Fig.\ \ref{slices} we show 3-d plots of
$\rho(x)$ over one of the slices and follow the evolution of the lump as we
apply more and more filtering (top to bottom). The two columns on the
left-hand side are for smearing, while the two right-hand side columns are for
the corresponding cooled sequence. It is obvious from the plots
that the position of the lump remains unchanged as the configurations become 
smoother and smoother. For both methods 
the height of the lumps decreases for the smoother configurations, 
but their width is
not altered drastically. Cooling seems to have the tendency to remove spiky
structures faster than smearing, which is on the other hand obvious since such
spikes cost a lot of action which is systematically reduced by cooling. 
 
For some of our configurations we find that the position of the lump in the
zero mode changes when switching the temporal boundary condition from periodic
to anti-periodic. Often the mode is then located at a completely different site
of the lattice and the spot where $\rho(x)$ showed a peak before now has a
flat distribution of $\rho(x)$. Thus for some of our $|Q| = 1$ configurations
we find two ``hot spots'' where the lump in the zero mode density is
located. In Fig.\ \ref{slices} we use a configuration which shows hopping and
show the slices through the respective maxima for both boundary conditions as
indicated in the header for each column. 
We find that both, cooling and smearing, detect the same hot spots and thus
also in this respect behave essentially identical. 

We found that for some configurations the hopping can stop after a
certain amount of smoothening and the zero mode settles at a single
position. Also for such cases we established that cooling and smearing behave
essentially identically, and for 5 of our 6 $|Q| = 1$ configurations we found that
the pattern for the hot spots agrees between the cooled sequence and its
smeared counterpart.    

\begin{figure}[t]
\centering
\includegraphics[width=11cm,clip]{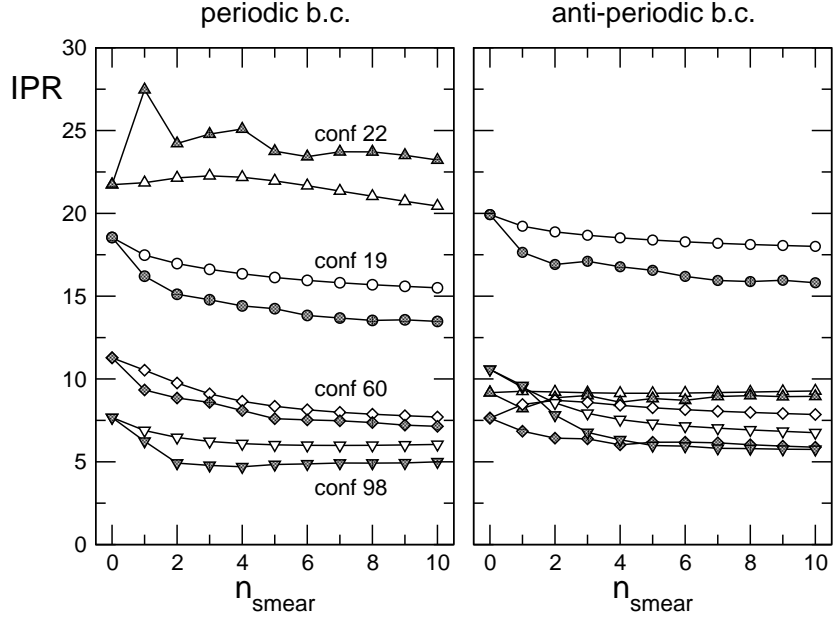}
\caption{The inverse participation ratio of zero modes as a function of the
  number of smearing (cooling) steps. The open symbols are for smearing, while
  the filled symbols represent the results from configurations with an
  equivalent number of cooling steps (compare Table 1). The left-hand side
  plot is for periodic boundary conditions, the right-hand side plot for
  the anti-periodic case. \hfill $^{\;}$}  
\label{iprcompare}
\end{figure} 

Let us finally come to the comparison of the behavior of the inverse
participation ratio $IPR$ under our two smoothing methods. In Fig.\
\ref{iprcompare} we show the history of IPR under cooling (filled symbols) and
smearing (open symbols). On the horizontal axis for simplicity we only display 
the number of smearing steps. The number of cooling steps, giving rise to 
configurations with equivalent action, is as listed in Table 1. The left-hand
side plot is for periodic temporal boundary conditions, the right-hand side is
for the anti-periodic case. 

For all cases cooling and smearing give rise to similar histories for the
inverse participation ratio. The values of $IPR$ remain in the same range,
although cooling typically produces somewhat smaller numbers (except for one
case). This is in agreement with the above observation, that cooling tends to
cut spiky structures with their generically higher inverse participation ratio. 

\section{Summary and conclusions}

In this contribution we present a comparison of smearing and cooling for the
removal of short range 
ultraviolet fluctuations from quenched SU(2) configurations. The
observables we use for this comparison are entirely based on the eigensystem
of the Dirac operator. The low lying Dirac eigenmodes couple to the long range 
structures of the gluon field and allow to study them without any 
manipulation of the gauge configuration. Thus the fermionic
observables are very suitable tools for comparing the different methods of
smoothing. 

We find that the two methods we apply, cooling and smearing, lead to
surprisingly similar results: Patterns in the eigenvalue spectrum,
characteristic for individual configurations are conserved to a large degree
by both smoothing methods. Furthermore, the topological charge as determined
by the index theorem remains unchanged in the range of smoothing we
consider. When analyzing the eigenvectors, we find that the shapes of lumps in
the scalar density change in the same way for cooling and smearing: The lumps
essentially only decrease in height, but remain at their position. Also the hopping
patterns of the zero modes under a change of the boundary conditions are essentially 
the same when applying cooling or smearing. A similar observation holds for
the inverse participation ratio which expresses the localization properties of
eigenmodes as a single number.

How should our findings be interpreted? We have motivated this study by the
quest for the perfect filter for long range structures in lattice gauge 
configurations. Our results seem to indicate that, at least for fermionic
observables, the two filtering methods applied, cooling and smearing, lead to a
relatively similar outcome. This could imply that the smoothing methods are
maybe less arbitrary than they seem at first glance and indeed leave 
the long range structures of the original configuration relatively
unchanged, at least for the amount of smoothing considered here. 
Certainly, this interpretation needs to be further tested by
analyzing more fermionic observables, as well as different gauge actions and the
gauge group SU(3). Also the evaluation of gluonic
observables on the sequences of configurations should be attempted. Both these
issues will be addressed in an upcoming study.    

\section*{Acknowledgments}
We thank Falk Bruckmann, Christian Lang, Michael M\"uller-Preussker, Andreas
Sch\"afer and Pierre van Baal for interesting
discussions. This work is supported by the DFG Forschergruppe {\sl Gitter
Hadronen Ph\"anomenologie}. The numerical calculations were done on the
Hitachi SR8000 at the Leibniz Rechenzentrum in Munich. We thank the LRZ staff 
for training and support.


\end{document}